\pdfoutput=1
\documentclass[aps,preprintnumbers,twocolumn,superscriptaddress,nofootinbib]{revtex4-1}
\usepackage{scrextend}
\usepackage{relsize}
\usepackage{amsmath}
\usepackage{amssymb}
\usepackage{epsfig}
\usepackage{graphicx}
\usepackage{hyperref}
\usepackage{dcolumn}   
\usepackage{tabu}
\usepackage{multirow}
\usepackage{color}
\usepackage[normal]{subfigure}
\usepackage{rotating}
\usepackage[margin=0.9in,a4paper]{geometry}
\usepackage[table]{xcolor}
\usepackage{enumitem}
\usepackage[utf8]{inputenc}
\usepackage{colortbl}
\usepackage{array,multirow}
\usepackage[normalem]{ulem}
\usepackage{comment}
\definecolor{nicered}{rgb}{0.7,0.1,0.1}
\definecolor{nicegreen}{rgb}{0.1,0.5,0.1}
\definecolor{red}{rgb}{1.0, 0, 0}
\usepackage{mathtools}
\usepackage{slashed}

\def\Tr{\mbox{Tr}\,}

\def\gsim{\raise0.3ex\hbox{$\;>$\kern-0.75em\raise-1.1ex\hbox{$\sim\;$}}}
\def\lsim{\raise0.3ex\hbox{$\;<$\kern-0.75em\raise-1.1ex\hbox{$\sim\;$}}}

\def\mb[#1]{\mathbf{#1}}

\renewcommand{\bar}{\overline}

\definecolor{LightCyan}{rgb}{0.88,1,1}
\definecolor{piggypink}{rgb}{0.99, 0.87, 0.9}
\definecolor{applegreen}{rgb}{0.55, 0.71, 0.0}
\definecolor{darkpastelgreen}{rgb}{0.01, 0.75, 0.24}
\definecolor{green-yellow}{rgb}{0.68, 1.0, 0.18}
\definecolor{darkgreen}{rgb}{0.0, 0.6, 0.0}

\newcommand{\beq}{\begin{equation}}
\newcommand{\eeq}{\end{equation}}
\newcommand{\beqa}{\begin{eqnarray}}
\newcommand{\eeqa}{\end{eqnarray}}

\newcommand{\be}{\begin{equation}}
\newcommand{\ee}{\end{equation}}
\newcommand{\bea}{\begin{eqnarray}}
\newcommand{\eea}{\end{eqnarray}}
\newcommand{\bi}{\begin{itemize}}
\newcommand{\ei}{\end{itemize}}
\makeatletter
\newcommand\notsotiny{\@setfontsize\notsotiny\@vipt\@viipt}
\makeatother

\newcommand{\IR}{\textnormal{\notsotiny \textsc{IR}}}
\newcommand{\UV}{\textnormal{\notsotiny \textsc{UV}}}
\newcommand{\A}{\textnormal{\notsotiny \textsc{A}}}

\newcommand{\I}{\textnormal{\notsotiny \textsc{I}}}
\newcommand{\G}{\textnormal{\notsotiny \textsc{G}}}

\newcommand{\B}{\textnormal{\notsotiny \textsc{B}}}

\newcommand{\AddrMainz}{%
PRISMA$^{+}$ Cluster of Excellence \& Mainz Institute for Physics, Johannes Gutenberg-Universit\"at Mainz, 55099 Mainz, Germany
}

\begin{document}


\title{The Super-Planckian Axion Strikes Back}

\author{Nayara~Fonseca}
\email{nayara.fonseca@desy.de}
\affiliation{\normalsize\it DESY, Notkestrasse 85, 22607 Hamburg, Germany}
\author{Benedict~von~Harling}
\email{bvonharling@ifae.es}
\affiliation{\normalsize\it DESY, Notkestrasse 85, 22607 Hamburg, Germany}
\affiliation{\normalsize\it IFAE, Universitat Aut\` onoma de Barcelona, 08193 Bellaterra, Barcelona, Spain}
\author{Leonardo~de~Lima}
\email{leonardo.de.lima@uffs.edu.br}
\affiliation{\normalsize\it Universidade Federal da Fronteira Sul, Av. Edmundo Gaievski 1000, 85770-000 Realeza, Brazil}
\author{Camila S. Machado}
\email{camachad@uni-mainz.de}
\affiliation{\normalsize\it \AddrMainz}

\begin{abstract}
  \noindent
We present a novel framework for obtaining large hierarchies in axion decay constants as well as trans-Planckian field excursions, with no need for tuning or a large number of fields. We consider a model with two or more CFTs with a common cutoff, that are linked by a gauged diagonal symmetry. This construction is dual to the geometry of a warped space with two or more throats glued at a common brane, allowing for calculability. Many applications of our setup are possible, such as ultra-light axions, natural inflation, and relaxion models.
\end{abstract}

\preprint{DESY 19-063}

\maketitle

\section{Introduction}

Axions are present in
several well-motivated extensions of the Standard Model.
These include the QCD axion to solve the strong $CP$-problem \cite{Peccei:1977hh, Wilczek:1977pj,Weinberg:1977ma}, axion inflation \cite{Freese:1990rb, ArkaniHamed:2003wu}, the relaxation mechanism to solve the hierarchy problem  \cite{Graham:2015cka}, ultra-light dark matter and dark energy \cite{Hlozek:2014lca, Kim:2015yna, Hui:2016ltb, Kobayashi:2018nzh}, and models of tachyonic particle production \cite{Anber:2009ua} (see also \cite{Hook:2016mqo, Agrawal:2017eqm, Machado:2018nqk}). Moreover, axion-like particles are abundant in string compactifications \cite{Marchesano:2014mla, Blumenhagen:2014gta,Hebecker:2014eua}. For these reasons, among others,  axion searches using a variety
of techniques are a very active field, see e.g. \cite{Irastorza:2018dyq}.

The axion has a sinusoidal potential generated by non-perturbative gauge configurations, which are responsible for the breaking of the continuous  shift symmetry to a remnant discrete one.  The leading potential for an axion $a'$ respecting the residual symmetry $a' \rightarrow a' + 2\pi f'$ can be written as
\begin{align}
  V(a') \sim &\, \Lambda^4 \left[ 1- \cos\left(\frac{a'}{f'}\right)\right]\,,
  \label{AxionPotential}
\end{align}
where $f'$ is the axion decay constant and $\Lambda$ is the scale associated with the non-perturbative physics. Since the shift symmetry protects their potential against large corrections,  axion-like particles are especially suitable for models requiring large field excursions.

Although several applications require large decay constants, there are many obstacles in finding consistent UV completions that generate $f'> M_{\rm Pl}$, where $M_{\rm Pl}$ is the (reduced) Planck mass. In particular, it has proven difficult to obtain axions with super-Planckian decay constants directly from String Theory \cite{Banks:2003sx,Baumann:2014nda,Bachlechner:2014gfa}. Furthermore, the Weak Gravity Conjecture (WGC) \cite{ArkaniHamed:2006dz} applied to $0$-forms would mean that higher-order corrections to the potential become important for such axions.

One way to circumvent some of these issues are models having two or more sub-Planckian axions at high energies which combine at low energies in a way that furnishes a super-Planckian axion.  A well-known construction along these lines is the Kim-Nilles-Peloso (KNP) mechanism \cite{Kim:2004rp} (see also \cite{Berg:2009tg, Ben-Dayan:2014zsa, Kappl:2014lra}), where this is achieved by a suitable alignment of the axion potentials. Although this is an attractive solution, it requires tuning of charges \cite{Choi:2014rja,Ben-Dayan:2014zsa}.  A  generalization of this mechanism is possible in a system with $N$ axions \cite{Choi:2014rja,Higaki:2014pja, Kaplan:2015fuy,Choi:2015fiu, Fonseca:2016eoo}. However, this in general requires a considerable number of fields in order to achieve large field excursions.
 Another possibility to obtain mild trans-Planckian  field values has been pointed out in \cite{Shiu:2015xda, Shiu:2015uva},  where two axions acting as St\"uckelberg fields are mixed through a gauge field.~\\

In this letter, we point out that an exponential hierarchy of decay constants can be naturally obtained if the axions arise as composite states from some (nearly) conformal sectors.  Our minimal model requires only two such sectors, charged under a weakly gauged abelian symmetry, which is broken when they undergo confinement at low energies.
Since the decay constants of the resulting St\"uckelberg fields are set by the respective confinement scales, in contrast to other constructions, they can easily be separated by a large hierarchy, with no tuning. This setup can be given a dual description by using the AdS/CFT correspondence in terms of slices of AdS$_5$ space glued together at a common brane  \cite{Cacciapaglia:2006tg} (see also \cite{Cacciapaglia:2005pa, Flacke:2006ad, Chen:2004gc, Chen:2005ad, Chialva:2005zy}), making it calculable. Our main result in this description then acquires a pleasing interpretation, where an exponential hierarchy of scales can be obtained uniquely from geometry. Furthermore, in order to explore the rich model building possibilities of this framework, we also study models with more than two conformal sectors, which are dual to multiple throats. We then discuss how axions with hierarchically different decay constants can be obtained using this setup.

\newpage
\section{Super-Planckian Decay Constants from CFTs/Throats} \label{sec:super2throats}
We consider two conformal sectors, each of them having a global $U(1)$ and $SU(N)$ symmetry. The diagonal subgroups of the $U(1)$s and $SU(N)$s are gauged by two vector fields $A_\mu$ and $G_\mu$, respectively. The Lagrangian reads\footnote{We follow the usual conventions $\mathrm{Tr} [T^a T^b] = \delta_{ab}/2$ and $\tilde{G}^{\mu\nu}=\epsilon^{\mu\nu\alpha\beta}G_{\alpha \beta}/2$.}
\begin{align}
&\mathcal{L} = \sum_{i=1}^2 \left( \mathcal{L}_{\rm CFT, i} + A_\mu J^\mu_{i} + \mathrm{Tr}\big[ G_\mu \mathcal{J}^{\mu}_i \big] \right) \nonumber \\
& - \frac{1}{4 g^2} F_{\mu\nu}F^{\mu\nu} - \frac{1}{2 g^{\prime 2}} \mathrm{Tr}\big[G_{\mu\nu}G^{\mu\nu}\big],
\end{align}
where $\mathcal{L}_{\rm CFT, i}$ are the Lagrangians of the two CFTs and $J^\mu_{i}$ and $\mathcal{J}^\mu_{i}$ are the $U(1)$ and $SU(N)$ currents, respectively. The traces are over the $SU(N)$ color indices. Next we assume that the CFTs have mixed $U(1)-SU(N)^2$ anomalies and that they undergo confinement at low energies. From anomaly matching (and since the $SU(N)$ symmetries are gauged), we know that the theory at energies below the confinement scales must contain composite scalars $a_i$ which encode these anomalies. Since also the $U(1)$ symmetries are gauged, we have under a gauge transformation
\begin{align}
&a_i(x)  \,\rightarrow a_i(x) + \Lambda(x) \,, \nonumber \\
&A_\mu(x)  \,\rightarrow A_\mu(x) - \partial_\mu \Lambda(x)\,.
\end{align}
The $a_i$ have the right quantum numbers to act as St\" uckelberg fields for $A_\mu$ and we expect the effective Lagrangian below the confinement scales to be
\begin{align}
&\mathcal{L}_{\rm eff} =  \sum_{i=1}^2 \left\{ \frac{f_i^2}{2} \left(A_{\mu} +\partial_\mu a_i \right)^{2} +\frac{c_i a_i}{8 \pi^2} \mathrm{Tr}\big[G_{\mu\nu} \tilde{G}^{\mu\nu}\big] \right\} \nonumber  \\
& - \frac{1}{4 g^2} F_{\mu\nu}F^{\mu\nu} - \frac{1}{2 g^{\prime 2}} \mathrm{Tr}\big[G_{\mu\nu}G^{\mu\nu} \big] + \dots \, ,
\label{EffectiveLagrangian}
\end{align}
where $f_i$ are the decay constants of the $a_i$ which are of the order of the respective confinement scales, the coefficients $c_i$ encode the mixed $U(1)-SU(N)^2$ anomalies, and the ellipsis stands for contributions from other composite states which we will take into account later.
One combination of scalars, which we will call $\tilde{a}$, is eaten, and the other combination $a'$  will be identified as the physical axion. These are given by
\begin{align}\label{e:basis}
\tilde{a} = \frac{f_1^2 a_1 +f_2^2 a_2}{\sqrt{f_1^2+f_2^2}}\,, \quad a' = \frac{f_1 f_2 (a_1-a_2)}{\sqrt{f_1^2+f_2^2}}\,,
\end{align}
where the new basis is defined such that $\tilde{a},~a'$ are canonically normalized. In terms of these fields, the Lagrangian can be written as
\begin{align}
&\mathcal{L}_{\rm eff} = \frac{f_1^2+f_2^2}{2} \left( A_{\mu} +\frac{ \partial_\mu \tilde{a}}{\sqrt{f_1^2+f_2^2}} \right)^{\!\!2} +\frac{1}{2} \left(\partial_\mu a' \right)^2 \nonumber \\
& +\frac{1}{8 \pi^2} \left(\frac{\tilde{a}}{\,\tilde{f}}+ \frac{a'}{f'}\,\right) \mathrm{Tr}\big[G_{\mu\nu} \tilde{G}^{\mu\nu}\big] \nonumber \\
& - \frac{1}{4 g^2} F_{\mu\nu}F^{\mu\nu} - \frac{1}{2 g^{\prime 2}} \mathrm{Tr}\big[G_{\mu\nu}G^{\mu\nu}\big] + \dots
\end{align}
The physical mass of the $U(1)$ gauge field is given by $M_A = g \sqrt{f_1^2+f_2^2}$. The effective decay constants of $\tilde {a}$ and $a'$ are
\begin{align}
\label{DecayConstants}
\tilde{f} = \frac{\sqrt{f_1^2+f_2^2}}{|c_1+c_2|}\,, \quad   f' = \frac{f_1 f_2 \sqrt{f_1^2+f_2^2}}{|c_1 f_2^2-c_2 f_1^2|}\,.
\end{align}
Under gauge shifts we now have
\begin{align}
&\tilde{a}(x)  \,\rightarrow \tilde{a}(x) + \Lambda(x) \sqrt{f_1^2+f_2^2}\,, \nonumber \\
 &A_\mu(x)  \,\rightarrow A_\mu(x) - \partial_\mu \Lambda(x)\,,
\end{align}
while $a'$ is neutral. For $c_1 + c_2 \neq 0$, this gauge transformation is anomalous due to the coupling to $\mathrm{Tr}[G_{\mu\nu} \tilde{G}^{\mu\nu}]$.
As in \cite{Shiu:2015xda}, we cancel this anomaly by adding a multiplet of chiral fermions which are external to the CFT and are charged under the gauge symmetries (see Appendix \ref{app:anomaly} for more details). We may now safely integrate out the massive gauge field $B_\mu \equiv A_{\mu} +\, \partial_\mu \tilde{a}/\sqrt{f_1^2+f_2^2}$, leaving only the physical axion $a'$ with decay constant $f'$.

Let us consider the case where only CFT$_1$ has a $U(1)-SU(N)^2$ anomaly, corresponding to $c_2=0$. Since the decay constants $f_1$ and $f_2$ are of the order of the confinement scales of the CFTs, a wide range of hierarchies among them is easily obtained. We in particular get:
\begin{equation}
\label{HierarchicalDecayConstants}
f' \simeq  \begin{dcases*}  \frac{f_1^2}{|c_1| f_2}\,, & $f_1 \gg f_2 $  \\
 \frac{f_1}{|c_1|}\,, & $f_1 \ll f_2 $\,.\end{dcases*}
\end{equation}
For example, if CFT$_1$ confines already near the Planck scale, $f_1\sim M_{\rm Pl}$, we easily obtain a super-Planckian decay constant for $f_2 \ll M_{\rm Pl}$.
The possibility to enhance the decay constant to trans-Planckian values through a mixing induced by gauge fields was already pointed out in Refs.~\cite{Shiu:2015xda,Shiu:2015uva}.
However, only a modest enhancement $\mathcal{O}(10 M_{\rm Pl})$ was considered there. In our construction, the decay constants are generated through dimensional transmutation such that an exponentially enhanced ratio $f_1/f_2$ can be generated without tuning charges as in alignment models \cite{Kim:2004rp, Ben-Dayan:2014zsa} or using a large number of axion fields \cite{Choi:2014rja,Higaki:2014pja} such  as in clockwork models \cite{Kaplan:2015fuy,Choi:2015fiu} or the $N$-relaxion \cite{Fonseca:2016eoo}. This exponential enhancement will be immediately obvious to see in the dual picture presented below.

\begin{figure}
 \center
\includegraphics[width=.47\textwidth]{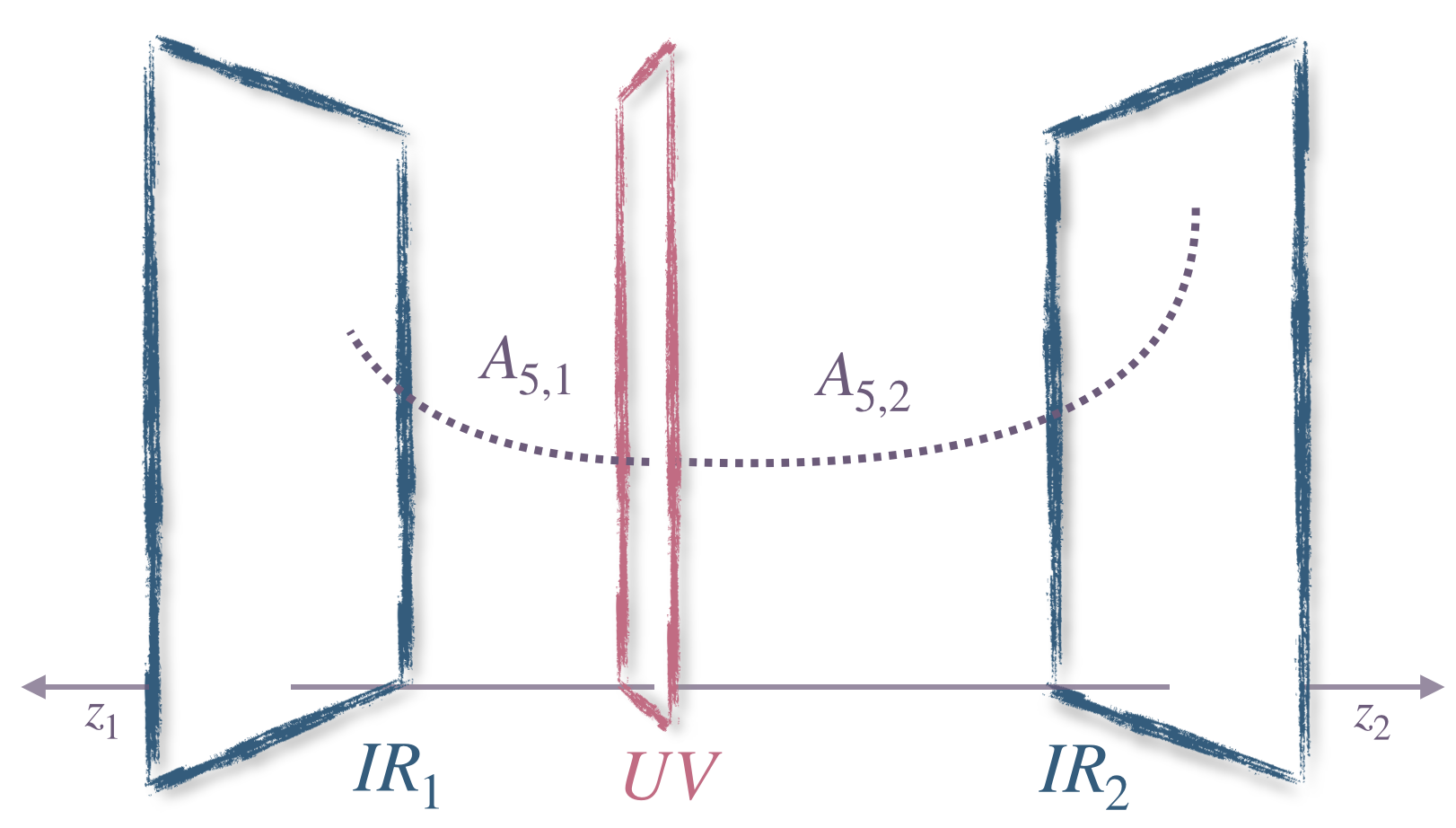}
\caption{Schematic drawing of the two-throat construction. The coordinates $z_1$ and $z_2$ grow from the UV brane to the IR$_1$ and IR$_2$ branes, respectively. The wave-functions of the fields $A_{5,1}$ and $A_{5,2}$ are represented by dashed lines.}
\label{fig:2throats}
\end{figure}

In order to confirm the above reasoning and to take into account the composite states that we have neglected, we next consider the dual perspective and assume that each CFT can be described by a slice of AdS$_5$ space \cite{Maldacena:1997re,ArkaniHamed:2000ds}, which are glued together at a common UV brane \cite{Cacciapaglia:2005pa,Cacciapaglia:2006tg} (cf.~Fig.~\ref{fig:2throats}). Each throat has proper length $L_i$ and, for simplicity, the same curvature scale $k$. In conformal coordinates the metric in each throat is
\begin{align}
\label{e:metric}
ds^2 \, = \, (k z_i)^{-2}\,(\eta_{\mu\nu} dx^\mu dx^\nu \, - \, dz_i^2) \,.
\end{align}
The warp factor is given by $(kz_i)^{-1}$,  with the UV brane at $z_\UV= 1/k$ and the IR branes at $z_{\IR_i}= e^{kL_i}/k$.
The stabilization of the two throats can be ensured by the Goldberger-Wise mechanism, analogously to the implementation in the Randall-Sundrum scenario, see e.g.~Ref.~\cite{Law:2010pv}.

The gauged $U(1)$ and $SU(N)$ symmetries of the CFTs are dual to corresponding gauge fields which propagate in the bulk of both throats. For later convenience, we consider separate $U(1)$ gauge fields $A_{M,i}$ in the two throats and then break the gauge symmetries to the diagonal subgroup on the UV brane. The $SU(N)$ gauge field $G_M$, on the other hand, is taken to propagate in both throats from the outset. Furthermore, the $U(1)-SU(N)^2$ anomalies are encoded by  mixed Chern-Simons terms in the bulk. The Lagrangian then reads 
\begin{align}\label{e:Sb}
&\sqrt{g}\,\mathcal{L}_{\rm B} = \!\sum_{i=1}^{2}\!\Bigg\{
 \frac{c_i}{16 \pi^2} \epsilon^{MNPQR}A_{M,i} \mathrm{Tr}\big[G_{NP} G_{QR}\big] \nonumber\\
&\!-\!\frac{(k z_i)^{-5}\!\!\!\!\!\!}{4 g_{5,i}^2}  F_{MN,i} F^{MN}_i\!\!-\!\frac{(k z_i)^{-5}\!\!\!\!\!\!}{2 g_{5}^{\prime 2}}~\mathrm{Tr}\!\left[G_{MN} G^{MN}\right]\!\!  \Bigg\}.
\end{align}
For simplicity, we will take the gauge couplings $g_{5,i}$ to be equal. The coefficients $c_i$ are the anomaly coefficients, taken to be integers.
At the UV brane, we consider
\begin{equation}\label{e:piuv}
\mathcal{L}_{\textrm{UV}} = \frac{v^2}{2} \left(\partial_\mu \pi_1 - A_{\mu,1} +A_{\mu,2} \right)^2 \,,
\end{equation}
where $\pi_1$ is the Goldstone mode of a bifundamental scalar and $v$ is its vacuum expectation value. We introduce $R_\xi$ gauge-fixing terms in the bulk and on the IR branes to decouple the vector and scalar modes and take the limit $\xi\rightarrow \infty$ \cite{Flacke:2006ad,Contino:2003ve}. 
The boundary conditions at the IR branes are then given by $A_{\mu,i}\vert_{\IR_i} = \partial_{z_i}(A_{5,i}/(k z_i))\vert_{\IR_i} =0$ and $\partial_{z_i} G_{\mu}\vert_{\IR_i} = G_{5} \vert_{\IR_i}=0$. Each abelian sector furnishes a scalar zero mode, a linear combination of which will be our axion, and the non-abelian sector has an unbroken gauge symmetry which will be responsible for the axion potential. Note that we do not introduce gauge-fixing terms for the abelian gauge bosons on the UV brane. Their vector and scalar modes thus still mix on the UV brane and we are not yet in unitary gauge.

Let us consider the holographic effective action at the UV brane. This is obtained by integrating out the bulk using profiles that satisfy the bulk equations of motion as well as the IR boundary conditions \cite{Panico:2007qd}. We will work at low energies, at lowest order in $p^2$ (see Appendix \ref{app:holo}),  or equivalently, neglecting all but the lowest mode in the Kaluza-Klein (KK) expansion (see Appendix \ref{app:kk}). In addition, for the non-abelian gauge boson, only the zero-mode contributes to the axion potential \cite{Grzadkowski:2007xm}.  The $p=0$ profiles for the abelian gauge fields are given by
\begin{align}
\label{e:wavefunction}
& A_{5,i} = k z_i \, a_i(x)\,,~~~A_{\mu,i}  = \frac{z_i^2-z_{\IR_i}^2}{z_{\UV}^2-z_{\IR_i}^2}\!A_{\mu,i}(x)\,,
\end{align}
and simply $G_5(x,z_i)=0$ and $G_{\mu}(x,z_i) = G_\mu(x)$ for the non-abelian gauge field. For future convenience, we have normalized the vector wave functions to unity on the UV brane. If there is no risk of confusion, the bulk fields and their respective $z_i$-independent zero modes are denoted by the same symbol. Plugging in these profiles and integrating over $z_i$, we obtain the effective Lagrangian
\begin{align}\label{e:Luv}
&\mathcal{L}_{\rm eff} = \frac{v^2}{2}\! \left(\partial_\mu \pi_1 - A_{\mu,1} +A_{\mu,2} \right)^2 \!-\!\frac{1}{2g^{\prime 2}} \mathrm{Tr}\big[G_{\mu\nu}G^{\mu\nu}\big] \nonumber \\
&+ \sum_{i=1}^2\bigg\{ \frac{1}{k g_5^2 \Delta z_i^2} \left(A_{\mu,i} + \frac{k \, \Delta z_i^2}{2} \partial_\mu a_i \right)^{\!2} \nonumber  \\
& + \frac{c_i k \, \Delta z_i^2}{16 \pi^2}\, a_i \mathrm{Tr}\big[G_{\mu\nu} \tilde{G}^{\mu\nu}\big]  -\frac{\mathcal{B}_i}{4g_5^2} F_{\mu\nu, i}F^{\mu\nu}_i \bigg\}\,,
\end{align}
where
\begin{align}\label{eq:Deltazi2}
\Delta z_i^2 \equiv z_{\IR_i}^2-z_{\UV}^2, \, \, \mathcal{B}_i = L_i \left[1 -\frac{3}{4 k L_i} + \mathcal{O}\left(\frac{z^2_{\UV}}{z^2_{\IR_i}}\right)\right]
\end{align}
and $g^{\prime -2} \equiv g_5^{\prime -2}( L_1+L_2)$. Taking the limit $v \rightarrow \infty$, we integrate out the bifundamental scalar and set $A_{\mu,1}=A_{\mu,2} \equiv A_{\mu}$. We now read off the effective coupling constant of $A_\mu$ to be $g^{-2} \equiv g_5^{-2}( \mathcal{B}_1+\mathcal{B}_2)$. We then rescale $a_i \rightarrow 2 a_i /(k \Delta z_i^2) $ which upon defining $f_i^2 \equiv 2 /(k g_5^2  \Delta z_i^2)$ reproduces Eq.~\eqref{EffectiveLagrangian} as expected due to the AdS/CFT duality.
Replacing the values of the physical parameters, at lowest order in $kL_i$ and $z_{\UV}^2/z_{\IR_i}^2$ we get the decay constant
\begin{equation}\label{eq:feff}
f'= \sqrt{\frac{2 k \,g^{-2}}{L_1+L_2}} \frac{\sqrt{e^{2 k L_1}+e^{2 k L_2}}}{|c_1 e^{2 k L_1}-c_2 e^{2 k L_2}|}\,.
\end{equation}
Taking the anomaly coefficient $c_2 =0$, we obtain
\begin{equation}
\label{e:feffwarped}
f' \simeq \begin{dcases*}
\sqrt{\frac{2k\, g^{-2}}{L_1+L_2}}\,\frac{e^{k (L_2-2 L_1)}}{|c_1|}\,, & $L_2 > L_1$ \\
\sqrt{\frac{2k \,g^{-2}}{L_1+L_2}}\, \frac{e^{- k L_1}}{|c_1|}\,, & $ L_1 > L_2$\,.
\end{dcases*}
\end{equation}
These results for the decay constant correspond to Eqs.~\eqref{DecayConstants} and \eqref{HierarchicalDecayConstants} in the dual description. As discussed in Appendix~\ref{app:kk}, they can also be derived using a KK expansion in the two-throat system.
The exponential enhancement for $L_2> L_1$ may be intuitively understood by noting that, for $c_2=0$, we have limited the anomalous coupling to only throat$_1$ while the axion $a'$ can propagate in the full space. This mismatch leads to a difference in the normalization between the axion kinetic term, which contributes to the numerator of Eq.~(\ref{eq:feff}) and the anomalous couplings, which contribute to the denominator, leading directly to the factor $\exp[k (L_2-2 L_1)]$ above (cf.~also Fig.~\ref{fig:2throats}).  As an illustration of the enhancement, for $k=10^{18}\, \mathrm{GeV}, \smash{c_1 \sim g\sim\mathcal{O}(1)}, L_2=20 \, k^{-1}, L_1=L_2/4$, we get $f'\sim 10^{22}\, \mathrm{GeV}$.

A common concern when faced with trans-Planckian axions is the WGC. In particular, important constraints arise from the coupling of the axion with gravitational instantons, see e.g.~\cite{Montero:2015ofa, Hebecker:2015zss, Brown:2015iha, Brown:2015lia}. The effective decay constant for this coupling arises from an integral over both throats.
Since the graviton and the axion propagate in all throats, in contrast to the case leading to Eq.~\eqref{e:feffwarped} there is no mismatch between the normalization of the axion kinetic term and its coupling to gravity which could lead to a super-Planckian decay constant. This implies that the axion couples to gravity with some decay constant $f_g \lesssim M_{\rm Pl}$ so that gravitational instantons satisfying the action bound $S_{\rm inst} \lesssim M_{\rm Pl}/f_g$ will not lead to large corrections to the axion potential \cite{Shiu:2018wzf}.


In order to generate a potential as in Eq.~\eqref{AxionPotential},
we assume that the $SU(N)$ gauge symmetry confines at a scale smaller than $f_{1,2}$. In Refs.~\cite{Shiu:2018unx,Shiu:2018wzf} (also considered in \cite{Shiu:2015xda, Shiu:2015uva}), it is argued that a non-vanishing potential for the axion field can be generated even though the fermions which cancel the anomalies are massless in the UV.  The resulting potential in  Refs.~\cite{Shiu:2018unx,Shiu:2018wzf} relies on the combination of the `t Hooft determinant term and four-fermion couplings which arise from integrating out the gauge field. Although this particular construction seems to differ from the literature \cite{Georgi:1981be, PhysRevLett.61.794, Banks:1994yg},\footnote{On the other hand,  one may argue that the limit $m_{u}\rightarrow 0$ cannot be unambiguously defined \cite{Creutz:2003xc} (see, however, also \cite{Srednicki:2005wc,slides}).} which agrees that the QCD  $\theta$ parameter is unphysical in the presence of massless quarks in the SM, the authors remark that the crucial difference in their mechanism is that there is only one generation of chiral fermions \cite{private}.

Independently of their construction, it is conceivable that a non-vanishing axion potential can be obtained if one considers some additional model building which gives masses to the fermions from an external source. For example, one can promote the fermions on the UV brane to bulk fermions and set the boundary conditions such that each one has a zero-mode. The latter contribute to the anomalies on the UV brane and allow to cancel them along the lines of Appendix~\ref{app:anomaly}. The fermions are in particular charged under the abelian gauge symmetry, forbidding mass terms in the bulk and on the UV brane. This symmetry is broken on the IR branes, on the other hand, and we can thus add mass terms for the fermions on these branes. If the zero-modes are localized towards the
UV brane, their resulting masses can be suppressed compared to the IR scales, allowing for a controllable size of the axion potential.\footnote{The bulk fermions also contribute to the Chern-Simons terms, see e.g.~\cite{Gripaios:2007tk}. If their bulk masses are somewhat larger than the AdS scale, as required for localized zero-modes, any perturbative corrections to the axion potential are highly suppressed, see e.g.~\cite{Pilo:2003gu,Contino:2003ve}.} 

In addition to obtaining a single trans-Planckian decay constant, using this  framework one can construct models that have multiple axion fields with hierarchically different decay constants. For  illustration,  let us double the spectrum such that we have two abelian fields $A_{M,i}$ and $B_{M,i}$ which can propagate in both throats with $i$ being the throat label. We also add two non-abelian gauge fields $G^\A_M$ and $G^\B_M$ which have  anomalous couplings to the $A_{M,i}$ and $B_{M,i}$, respectively. Let us limit the anomalous coupling  to $G^\A_M$ only to throat $i=1$  and the one to $G^\B_M$ only to throat $i=2$. Assuming $L_2>L_1$, we then have from Eq.~(\ref{eq:feff}) that the decay constants are exponentially separated
\begin{align}
 & \frac{f'_\A}{ f'_\B} \, \approx \, e^{2k(L_2-  L_1)} \, \gg \,1\,,
\end{align}
where, for simplicity, we have taken the anomaly coefficients to be order one. The potential at low energies can then be written as
\begin{align}
 \! V \!= \!\Lambda_{\A}^4\left[1 - \cos \!\left(\frac{ a'}{f_{ \A}^{\prime} }\right) \right] \!+ \Lambda_{\B}^4 \left[1 -\cos \!\left(\frac{ b' }{f_{\B}^{\prime}}\right)\right],
\end{align}
where $a'$ and $b'$ refer to the uneaten 4d scalar fields. This setup can, for example, be applied to a two-component axion model with an ultra-light axion, which can constitute most of the dark matter, and the QCD axion as the other component  \cite{Kim:2015yna}. Such models have clear phenomenological consequences as one can  explicitly compute all the couplings of the axion-like fields to the Standard Model particles. The details of this construction will be presented in a future work.

Another possibility is to consider only the field $A_{M,i}$ coupled to both $G^{\A,\B}_M$, to generate a single field potential with two cosines with hierarchically different decay constants, as required in relaxion models \cite{Graham:2015cka}. Such a construction was explored in \cite{Fonseca:2017crh}.

\section{Several CFTs/throats: a playground for generating hierarchies}\label{sec:multi}

Let us generalise the setup of the previous section to several CFTs with a common cutoff, which are then dual to several throats with a common UV brane. Such a multi-throat setup combined with kinetic mixing and alignment allows us to construct scenarios with multiple hierarchical decay constants.  In particular, we can reproduce the alignment mechanism with two cosines in the potential, as in the original KNP model, and also obtain the alignment with $N$ axions, as in clockwork models. Moreover, the multi-throat construction, with warped (or also flat) geometry, provides many possibilities for model building.

We add one $U(1)$ gauge field  $A_{M,i}$ with gauge coupling $g_{5,i}$ for each throat, where $i=1,...\,,N$ and $N$ is the number of throats. In order to break the $U(1)$ gauge symmetries on the IR branes, we then impose the boundary conditions $A_{\mu,i}\vert_{\IR_i}=\partial_{z_i}\left(A_{5,i}/(k z_i) \right)\vert_{\IR_i}  =  0$.

As in Sec.~\ref{sec:super2throats}, the UV brane contains bifundamental scalars $\pi_i$ linking $A_{\mu, i}$ and $A_{\mu,i+1}$. For simplicity, we will work directly in the limit where $v_i\rightarrow \infty$, such that only the diagonal gauge symmetry survives, which amounts to identifying the vector fields at the UV brane, i.e.~$A_{\mu,i}\vert_{\UV} \equiv A_\mu$. Furthermore, of the $N$ scalars coming from the scalar zero modes, one linear combination will be eaten by the diagonal vector field, leaving only $N-1$ propagating scalars in unitary gauge \cite{Cacciapaglia:2005pa,Cacciapaglia:2006tg}.

Next, we can add anomalous couplings such that each $A_{M,i}$ couples with different non-abelian gauge groups, each one characterized by a capital letter superscript ${\rm I}={1,\ldots, n_\G}$. Using the axion profiles $A_{5,i} = k z_i a_i(x)$, we then get  $\mathcal{L}_{\rm eff} \, \supset \, \sum_{\I=1}^{n_\G} \frac{k \,\Delta z_i^2}{2}\, \mathcal{C}_{i}^{~\!\!\I}$,  where
\begin{align}
\label{eq:cseffective}
&\mathcal{C}_{i}^{~\!\!\I} \, \equiv \, \frac{c_i^{~\!\!\I}}{8 \pi^2} a_i \mathrm{Tr} \left[ G^{~\!\!\I}_{\mu \nu} \tilde{G}^{~\!\!\I, \mu \nu} \right]\, ,
\end{align}
$\Delta z_i^2$ is given in Eq.~(\ref{eq:Deltazi2}), and $c_i^{~\!\!\I}$ are the anomaly coefficients. These couplings generically lead to an anomaly of the diagonal gauge symmetry at the UV brane, which may be cancelled by adding suitable fermion multiplets  \cite{Shiu:2015xda} or by imposing that the relation $\sum_{i,~\!\!\I} c_i^{~\!\!\I}=0$ is fulfilled (for details, see Appendix \ref{app:anomaly}).

Repeating the same steps as in Sec.~\ref{sec:super2throats}, the effective Lagrangian becomes
\begin{align} \label{eq:generalLagrangian}
&\mathcal{L}_{\rm eff} \, = \, \sum_{i=1}^N\left\{ \frac{f_i^2}{2} \left(A_{\mu} +\partial_\mu a_i \right)^2 +\sum_{\I=1}^{n_\G} \mathcal{C}_{i}^{~\!\!\I}\right\} \nonumber \\&- \frac{1}{4 g^2} F_{\mu\nu}F^{\mu\nu}- \sum_{\I=1}^{n_\G} \frac{1}{2 g^{\prime 2}_{~\!\!\I}} \mathrm{Tr}\left[G_{\mu\nu}^{~\!\!\I} G^{~\!\!\I,\mu\nu}\right]\,,
\end{align}
where $f_i^2 \equiv 2 /(k g_5^2  \Delta z_i^2)$ as before. The coupling to $A_\mu$ induces a mixing among the axion fields. At this point, one should perform an $SO(N)$ rotation to bring the fields from the $a_i$ basis to a new basis where one linear combination $\tilde{a} \propto \sum_{i=1}^N  f_i^2 \, a_i $ is eaten and $N-1$  modes remain. 

Using this setup for three throats, $N=3$, and a single non-abelian field, $n_\G=1$, we can obtain an enhanced effective decay constant with a one-cosine potential by imposing  a discrete $\mathbb{Z}_3$ symmetry under exchange of the throats.
Moreover, for the case $N=3$ and  $n_\G=2$, we reproduce a KNP-like alignment. Another possibility is to consider $N$ throats with $n_\G=N$ which leads to an alignment system with $N$ axions.  We illustrate these examples in Appendix \ref{app:examples}.

\section{Conclusions}
In order to achieve super-Planckian and more generally hierarchical decay constants for a system with multiple axions, we have considered a 4d theory with multiple CFTs with a common cutoff. The CFTs have global $U(1)$ symmetries, whose diagonal subgroup is weakly gauged. The symmetries are broken when the CFTs confine, leading to corresponding Goldstone bosons which mix through their coupling to the gauge boson. Each CFT has a distinct strong-coupling scale.
An exponential hierarchy among the decay constants is then achieved naturally through dimensional transmutation.

The dual picture corresponds
to a warped multi-throat geometry. An exponential enhancement of the decay constants is given by the warped factor without requiring a large number of axions or large charges. Specifically, the enhancement is controlled by the difference of the throat lengths and has a simple geometric interpretation as can be seen in Fig.~\ref{fig:2throats} and Eq.~(\ref{e:feffwarped}).

A possible direction for future investigation is whether a string embedding for our construction is attainable. In addition, the multi-throat scenario is an interesting playground for several applications. Using this setup, one can naturally obtain models where the dark matter is composed of multiple axions, constructions with interacting dark matter and dark energy, and realizations that require multiple hierarchies of decay constants, to mention a few examples. We hope that this framework can provide new model building avenues.

\subsection*{Acknowledgments}
We thank Sebastian Ellis, Arthur Hebecker, Rachel Houtz,  Oriol Pujolas, Fabrizio Rompineve,  Alexander Westphal, and Fang Ye for useful discussions. We also thank Gary Shiu and Wieland Staessens for correspondence
regarding their work. The work of NF was supported by the Deutsche Forschungsgemeinschaft under Germany's Excellence Strategy - EXC 2121 ``Quantum Universe" – 390833306. The work of CSM was supported
by the Alexander von Humboldt Foundation, in the framework of the Sofja Kovalevskaja Award 2016, endowed by the German Federal Ministry of Education and Research and also supported by  the  Cluster  of  Excellence  ``Precision  Physics,  Fundamental
Interactions, and Structure of Matter" (PRISMA$^+$ EXC 2118/1) funded by the German Research Foundation (DFG) within the German Excellence Strategy (Project ID 39083149).

\begin{center}
\scalebox{.04}{\underline{We also thank Gia Dvali for his simple and general arguments.}}
\end{center}

\appendix



\section{Gauge anomaly cancellation} \label{app:anomaly}

Let us consider the action for $N$ throats, with bulk couplings in each throat of the form
\be
\mathcal{L}_B \supset \frac{c_i^{~\!\!\I}}{16 \pi^2} \epsilon^{MNPQR}A_{M,i} \mathrm{Tr}\big[G^{~\!\!\I}_{NP} G^{~\!\!\I}_{QR}\big] \,,
\ee
where the index $i$ runs over the throats and the index ${\rm I}$ over the different $SU(N)$ groups. On the UV brane, we can also add couplings of the bifundamental fields $\pi_i$, which link $A_{\mu,i}$ to $A_{\mu,i+1}$, to the non-abelian groups:
\begin{align}
&\mathcal{L}_\UV  \supset \frac{c_{\pi, i}^{~\!\!\I}}{8\pi^2}\pi_i \,\Tr[G_{\mu \nu}^{~\!\!\I} \tilde{G}^{~\!\!\I,\mu \nu} ]\,.
\end{align}
With such terms present, a gauge transformation $A_{M,i}\rightarrow A_{M,i}-\partial_M\,\Lambda_i (x,z_i)$ generates brane-localized anomalies
\be
\delta \mathcal{L}_{\rm brane} =  \,  \sum_\I \frac{c_{\pi, i-1}^{~\!\!\I} -c_{\pi, i}^{~\!\!\I}\pm c_i^{~\!\!\I}}{8 \pi^2}  \Lambda_{i}(x)  \mathrm{Tr}\big[G_{\mu \nu}^{~\!\!\I} \tilde{G}^{~\!\!\I,\mu \nu} \big] ,
\label{BraneLocalizedAnomalies}
\ee
where $c^{~\!\!\I}_{\pi, 0}=c^{~\!\!\I}_{\pi, N}=0$, the $c^{~\!\!\I}_{\pi, i}$ are present only at the UV brane, the plus (minus) sign is obtained at the UV (IR$_i$) brane and $\Lambda_i(x)$ equals $\Lambda_i(x,z_i)$ at the respective brane.

Admitting the boundary conditions $A_{\mu,i}|_{\IR_i}=0$, for all $i$, the IR anomalies are global and hence harmless. The anomalies at the UV brane, on the other hand, are cancelled if for each $i$ and $\textrm{I}$ the condition $c_{\pi, i-1}^{~\!\!\I}-c_{\pi, i}^{~\!\!\I} +c_i^{~\!\!\I}=0$ is satisfied.

In addition, we can get contributions to cancel the anomalies by adding suitable fermions on the UV brane. For simplicity, let us focus on the case that the strong groups $SU(N)_\I$ propagate in at most two throats. For each strong group, we may add two pairs of chiral fermions, $\psi_L^{\alpha}$ and $\psi_{R}^{\alpha}$, where $\alpha = 1,2$ is a flavor index, which are charged under the $U(1)_i$ and which transform in the fundamentals and anti-fundamentals of the $SU(N)_\I$, respectively (see also \cite{Shiu:2015xda}). The fermionic Lagrangian on the UV brane is then
\begin{align}
\mathcal{L}_{\psi} = \sum_\alpha \bar{\psi}_L^{\alpha} \, i \sigma^\mu D_\mu \psi_L^\alpha+  \bar{\psi}_R^{\alpha} \, i \bar{\sigma}^\mu D_\mu \psi_R^\alpha \,,
\end{align}
where  $D_\mu = \partial_\mu + i \sum_i q^{\alpha,i}_{L,R} A_{i,\mu}+ i \, G_\mu^\I$ and $q^{\alpha,i}_{L,R}$ are the $U(1)_i$ charges of the fermions. Under a $U(1)_i$ gauge transformation
\begin{align}
 \psi_{L,R}^{\alpha} \rightarrow \psi_{L,R}^{\alpha}  \, e^{i \, q^{\alpha,i}_{L,R} \Lambda_i}\,,
\end{align}
the fermionic terms transform as
\begin{equation}
\!\!\delta\mathcal{L}_{\psi}\! =\! \sum_{i} \Lambda_i \left( \partial_{\mu} \mathcal{J}^{\mu}_{\psi, i}\! +\! \frac{\mathcal{A}_i^{\textrm{mix}}}{8 \pi^2} \, \Tr[G_{\mu \nu}^{~\!\!\I} \tilde{G}^{~\!\! \I,\mu\nu} ] \right),
\label{eq:fermiontransformation}
\end{equation}
where $\mathcal{A}_i^{\textrm{mix}}=\frac{1}{2}\sum_{\alpha}(q_L^{\alpha,i}-q_R^{\alpha,i})$ and $\mathcal{J}^{\mu}_{\psi, i}$ is the $U(1)_i$ current of the fermions. 

Let us focus on $N$=2 throats. The mixed $U(1)_i -\, SU(N)^2$ anomalies from the fermions in Eq.~\eqref{eq:fermiontransformation} cancel with those in Eq.~\eqref{BraneLocalizedAnomalies} and cubic $U(1)_i^3$ anomalies cancel among the fermions by choosing charges which satisfy
\begin{equation}
\!\!\!\mp c_{\pi, 1}^{~\!\!\I} +  \mathcal{A}_i^{\textrm{mix}}\! = \!-c_i^\I\,,~\sum_{\alpha}\!\left[(q_L^{\alpha,i})^3\!-\!(q_R^{\alpha,i})^3\right]\! = 0, \!
\end{equation}
where the minus (plus) sign refers to $i=1$ (${i=2}$). Adding the equations for the $i=1$ and ${i=2}$ mixed anomalies, one obtains the condition for the cancellation of the mixed anomaly of the diagonal gauge symmetry. Together with the corresponding condition for the cubic anomaly, these are the remaining constraints in the  $v \rightarrow \infty$ limit,
\begin{equation}\label{e:anomaly}
\sum_{\alpha}\!q_R^\alpha\!-\!q_L^\alpha\!=\!c_1^{~\!\!\I}\!+ c_2^{~\!\!\I}\,,~~\sum_{\alpha}\! (q_R^\alpha)^3\!-\!(q_L^\alpha)^3\!=\!0\,,
\end{equation}
where $q_{L,R}^\alpha \equiv \sum_i q_{L,R}^{\alpha,i}$.  These equations have solutions for integer charges when $c_1^{~\!\!\I}+ c_2^{~\!\!\I}$ is divisible by six. One possible choice, for the case $c_1^{~\!\!\I}+ c_2^{~\!\!\I}=6$, is given by $q_L^{1} = 12,~ q_L^{2}=1, q_R^{1} = 10,~ q_R^{2}=9$. For other possibilities, see \cite{Shiu:2015xda}. In particular, the inclusion of bulk Chern-Simons terms for the abelian gauge fields can alleviate the constraints on the charges and the $c_i^{~\!\!\I}$ that are needed in order to satisfy Eq.~(\ref{e:anomaly}), without modifying the axion potential. This also allows to satisfy the anomaly cancellation conditions with only one generation of fermions, $\psi_L$ and $\psi_{R}$ (cf.~the discussion about the axion potential in Sec.~\ref{sec:super2throats}).

\section{Holographic action and higher-dimensional operators}\label{app:holo}
Here we give the details of the procedure to obtain the effective Lagrangian in Eq.~(\ref{e:Luv}). Starting from the action in Eq.~(\ref{e:Sb}), the equations of motion for $A_{\mu,i}$ and $A_{5,i}$ in $R_\xi$ gauge read:\footnote{For holographic calculations, it is customary to work in the $A_5=0$ gauge. We prefer the $R_\xi$ gauge to avoid subtleties in the treatment of the topological terms.}
\begin{align}
&\left[\eta^{\mu\nu}\!\left(\Box- \frac{\partial_{z_i} \left(k z_i \partial_{z_i}\right)}{k z_i}\right)\!-\partial^\mu \partial^\nu \!\left(\!1-\frac{1}{\xi_i}\right)\right]\!A_{\nu,i}=0,\nonumber \\
&\,\Big[\Box +\xi_i \partial_{z_i} \left((k z_i)^{-1} \partial_{z_i}\right)\Big]\frac{A_{5,i}}{k z_i}=0.
\end{align}
We look for solutions of these equations of motion that satisfy the IR boundary conditions $A_{\mu,i}\vert_{\IR_i} =\partial_{z_i}(k z_i A_{5,i})\vert_{\IR_i}=0$. For simplicity, let us take the unitary gauge $\xi_i\rightarrow\infty$ in the bulk. This decouples all but the zero-momentum mode of $A_{5,i}$, such that the solution is $A_{5,i} = k z_i a_i(x)$, which is the same as in Eq.~(\ref{e:wavefunction}). It is also convenient to separate the vector into transverse and longitudinal polarizations $A_{\mu,i} = A^T_{\mu,i}+A^L_{\mu,i}$, such that $\partial^\mu A^T_{\mu,i}=0$. Going to momentum space, we make the ansatz $A^{T,L}_{\mu,i} (p,z) =f^{T,L}(p,z) A^{T,L}_{\mu,i} (p)$. The holographic profiles $f^{T,L}(p,z)$ are given by \cite{Panico:2007qd}
\begin{align}
&f^{T}\!(p,z) \!= \!\frac{z}{z_{\UV}}\frac{J_1(p z_{\IR_i}) Y_1(p z)\!-\!J_1(p z) Y_1(p z_{\IR_i})}{J_1(p z_{\IR_i}) Y_1(p z_{\UV})\!-\!J_1(p z_{\UV}) Y_1(p z_{\IR_i})\!} \, ,\nonumber \\
&f^{L}\!(p,z) \!=\! f^{T}(0,z)\! = \! \frac{z_i^2-z_{\IR_i}^2}{z_{\UV}^2-z_{\IR_i}^2},
\end{align}
and are normalized to unity on the UV brane. We note that at zero momentum, we recover the expression in Eq.~(\ref{e:wavefunction}). Inserting these profiles in the bulk and integrating, only the boundary terms at the UV brane survive, leading to the effective Lagrangian:
\begin{align}
&\mathcal{L}_{\rm eff} \supset  \sum_{i=1}^2 \left\{ \frac{1}{k g_5^2 \Delta z_i^2} \left(A_{\mu,i} + \frac{k \Delta z_i^2}{\!\!2} \partial_\mu a_i \right)^{\!\!2} \right. \nonumber \\
&\left. -\frac{L_i}{4 g_5^2} \left[1-\frac{3}{4 k L_i} +\mathcal{O} \left(\frac{z_{\UV}^2}{z_{\IR_i}^2}\right)\right] F_{\mu\nu, i}F^{\mu\nu}_i \right. \nonumber \\
&\left.+ \frac{7 z_{\IR_i}^2}{192 \, k g_5^2} \!\left[1+\mathcal{O} \left(\frac{z_{\UV}^2}{z_{\IR_i}^2}\right)\right]\! F_{\mu\nu, i}\Box F^{\mu\nu}_i+\cdots \!\right\}.
\end{align}
The coefficient of the kinetic term is identified as $\mathcal{B}_i$ in Eq.~(\ref{e:Luv}). We have also written the leading dimension-six correction. It is suppressed by $z_{\IR_i}^{-2} = k^2 \exp{(-2 k L_i)}$, which is of the order of the KK mass scale squared, as expected on general power counting grounds \cite{Chala:2017sjk}.

Similar considerations apply for the non-abelian sector, except for the different boundary conditions $\partial_{z_i} G_{\mu,i}\vert_{\IR_i} = G_{5,i} \vert_{\IR_i} =0$. For details about the higher-order interactions correcting the topological term, we refer the reader to Ref.~\cite{Grzadkowski:2007xm}.

\section{Kaluza-Klein derivation of the effective decay constant}
\label{app:kk}
Here we present an alternative derivation of the main results of Sec.~\ref{sec:super2throats} using a KK expansion in the two-throat system. We work in unitary gauge\footnote{Before fixing the gauge, the action is explicitly gauge invariant, as long as the anomaly cancellation conditions discussed in Appendix~\ref{app:anomaly} are satisfied.} and in the limit $v \rightarrow \infty$ such that there is one abelian vector field $A_M$ which propagates in all throats. The $U(1)$ gauge symmetry is broken on the IR branes by imposing the boundary conditions
\begin{align}
\label{e:bcs}
\left. A_{\mu,i}\right\vert_{\IR_i} \, =\,  0 \, , \qquad  \left. \partial_{z_i}\left( \frac{A_{5,i}}{k z_i} \right)\right\vert_{\IR_i} \, = \, 0 \,.
\end{align}
On the UV brane, the gauge fields need to satisfy the following boundary conditions in the limit $v \rightarrow \infty$ \cite{Cacciapaglia:2005pa,Cacciapaglia:2006tg}:
\begin{align}
\label{eq:UVthroats}
&\left(A_{\mu,i}-A_{\mu,i+1}\right)_{\UV}=0,\quad \sum_{i=1}^N \left. \frac{1}{k z_i g_{5}^2}\partial_{z_i} A_{\mu,i}\right\vert_{\UV}\!\!\!=0\,,\nonumber\\
&\sum_{i=1}^N \left. \frac{1}{k z_i g_{5}^2}A_{5,i}\right\vert_{\UV}=0\,,~ \left.\partial_{z_i} \left(\frac{A_{5,i}}{k z_i}\right)\right\vert_{\UV} =0\,.
 \end{align}
This arises from requiring that, on the UV brane,  the boundary terms vanish and the functions $A_{5,i}/(k z_i)$ and $\partial_{z_i} (A_{5,i}/(k z_i))$ are continuous. The boundary conditions allow for a massless mode for each throat with wavefunction
\be
\label{eq:A50}
A_{5,i} \, = \,\mathcal{N}_i \,k z_i \,a_i(x) \, ,
\ee
where the normalization constant $\mathcal{N}_i$ is obtained from demanding that the kinetic term of $a_i$ is canonically normalized. Then Eq.~(\ref{eq:UVthroats}) implies that
\begin{align}
\label{eq:cont}
\sum_{i=1}^N \mathcal{N}_i \, a_i =0\,.
\end{align}
We may use this equation to rewrite, for instance, the field $a_N$ in terms of the other $N-1$ 4d fields.  This is an important point in our construction as it leads to mixing in the axion moduli space.

Let us focus on the case with $N=2$ throats. Integrating Eq.~\eqref{e:Sb} over $z_i$, we get
\begin{align}
\label{e:kk4d}
S_{\rm{4D}}  \supset  \int d^4x \, \bigg\{ (\partial_{\mu} a_i)^2 +
\frac{c_i  a_i}{8 \pi^2 f_i}  \mathrm{Tr}\big[G_{\mu \nu} \tilde{G}^{\mu\nu}\big]\bigg\}
\end{align}
for $i=1,2$ and where \cite{Choi:2003wr,Flacke:2006ad}
\begin{align}
\label{eq:fB}
f_{ i}^{-1} &\equiv \int_{z_\UV}^{z_{\IR_i}} dz_i \, \mathcal{N}_i \,k z_i  =g_5 \sqrt{\frac{k\,\Delta z_i^2}{2}}
\end{align}
with $\Delta z_i^2$ given in Eq.~(\ref{eq:Deltazi2}). Assuming $k L \gg1$, we then get $f_{i}  \sim \, k\,e^{-k L_i} $. Now $a_1$ and $a_2$ are related by Eq. (\ref{eq:cont}). Therefore, we are left with just one degree of freedom with action
\begin{align}
S_{\rm{4D}}\!  \supset \! \!  \int d^4x  \,\bigg\{ (\partial_{\mu} a')^2 + \frac{a'}{8 \pi^2 f_{\rm eff}} \mathrm{Tr}\big[G_{\mu \nu} \tilde{G}^{\mu\nu}\big] \bigg\} \, ,
\end{align}
where $a' = \sqrt{1+\mathcal{N}_1^2/\mathcal{N}_2^2}\,a_1 $ and the effective decay constant $f_{\rm eff}$ is given by Eq.~(\ref{eq:feff}), reproducing our main result.

\section{Examples with $\boldsymbol{N>2}$ throats}
\label{app:examples}

In this appendix, we discuss three examples for a system with $N>2$ throats. Let us first explore the choice $N=3$ and $n_\G=2$ (with the non-abelian gauge fields $G^{\A,\B}_M$) which can be used to obtain a KNP-like alignment. This leads to a system of two axions which will allow us to obtain a potential that has an almost flat direction.
In order to read off the effective decay constant, we diagonalize the axion system by performing an $SO(3)$ rotation and then canonically normalize the fields. In the new basis, we have
\begin{align}
\label{e:newbasis}
&a'_1 = \frac{f_2(f_1^2 a_1-(f_1^2+f_3^2) a_2 +f_3^2 a_3)}{\sqrt{f_1^2+f_3^2}\sqrt{f_1^2+f_2^2+f_3^2}} \,, \\
&a'_2 = \frac{f_1 f_3 (a_1-a_3)}{\sqrt{f_1^2+f_3^2}}\,,~ \tilde{a} = \frac{f_1^2 a_1+f_2^2 a_2 +f_3^2 a_3}{\sqrt{f_1^2+f_2^2+f_3^2}} \nonumber \,.
\end{align}
Then the Lagrangian in Eq.~(\ref{eq:generalLagrangian}) becomes
\begin{align}
&\mathcal{L}_{\rm eff}^{\prime} = \frac{M_A^2}{2 g^2} \left( A_{\mu} +g \frac{ \partial_\mu \tilde{a}}{M_A} \right)^2 \!\!+\sum_{i=1}^2\frac{1}{2} \left(\partial_\mu a'_i \right)^2+\mathcal{L}_{\rm kin}\nonumber \\
&\!\!\!+\!\sum_{\I=\A,\B}\!\frac{1}{8 \pi^2} \left(\frac{\tilde{a}}{\,\tilde{f}_{~\!\!\I}}+\sum_{i=1}^2\frac{a'_i}{f_{~\!\!\I,i}^{\prime}}\,\right) \mathrm{Tr}\big[G^{~\!\!\I}_{\mu \nu} \tilde{G}^{~\!\!\I,\mu\nu}\big]\,,
\end{align}
where $\mathcal{L}_{\rm kin}$ are the gauge boson kinetic terms, $M_A = g \sqrt{f_1^2+f_2^2+f_3^2}$ is the gauge boson mass and $\tilde{f}_{~\!\!\I},~f_{~\!\!\I,i}^{\prime}$ are the axion decay constants which depend on the coefficients $c_{i}^{~\!\!\I}$ and on the decay constants $f_i$. We see that, in unitary gauge, $\tilde{a}$ is eaten by $A_\mu$ and disappears from the spectrum, while $a'_{1,2}$ are physical and uncharged under the diagonal $U(1)$.  The potential is then simply
\begin{align}
\label{e:align}
V &=  \sum_{I=A,B}\Lambda_{\I}^4\left[1- \cos\left(\frac{a_1'}{f_{\I,1}^{\prime}}+ \frac{a_2'}{f_{\I,2}^{\prime}}  \right)  \right] \,.
\end{align}
A super-Planckian decay constant can then be obtained by  appropriately choosing the anomaly coefficients, similarly to the KNP alignment mechanism \cite{Kim:2004rp}. However, due  to a mixing in the axion moduli space, there is a continuous parameter which can be used to alleviate the tuning on the anomaly coefficients \cite{Shiu:2015xda}. In this context, a trans-Planckian decay constant is disfavoured in the warped case, since the tuning of the anomaly coefficients or mixing angle has to compensate the exponential down-warping of the decay constants.  For a  flat metric, the tuning is just linear and one may still obtain a super-Planckian decay constant with reasonable parameters.

In another model building direction, we can increase the number of throats and obtain the alignment  for a system with many axions as in Ref.~\cite{Choi:2014rja}. This corresponds to the case with $N$ throats and $n_\G=N$ non-abelian gauge groups, which leads to $N^2$ anomalous couplings. The potential (before the $SO(N)$ rotation) is then
\begin{align}
V= \sum_{\I=1}^N \Lambda_{~\!\!\I}^4 \left[ 1- \cos\left(\sum_{i=1}^N\frac{  c^{~\!\!\I}_i\, a_i}{f_{~\!\!i}} \right) \right]\,.
\end{align}
In this case, it is possible to get a decay constant which is enhanced by a factor that goes as $\sim \sqrt{N!}\,c^{N-1}$, where $c$ denotes a typical value of the anomaly coefficients,  similar to the clockwork construction.

Another possibility is to consider $N=3$ throats and a single non-abelian gauge group, $n_\G=1$. In general, for a potential with just one cosine and multiple axions, $f_{\rm eff}$ is always sub-Planckian as $1/f_{\textrm{eff}}^{2} \equiv \sum_i 1/f_{\I,i}^{\prime 2}$. However, the non-trivial mixing from Eq.~(\ref{e:newbasis}) can lead to an enhancement as we show in the following.
At low energies, this example leads to just one of the terms of Eq.~(\ref{e:align}). The explicit form of $f_{\A,i}^\prime$ in this case is given by
\begin{align}
f_{\A,1}^\prime &= \frac{f_1 f_3 \sqrt{f_1^2+ f_3^2}}{ |c_3 f_1^2 - c_1 f_3^2|} \,,\nonumber\\
f_{\A,2}^\prime &=  \frac {f_2 \sqrt{f_1^2 + f_3^2(f_1^2+f_2^2+ f_3^2)}}{|c_2(f_1^2+ f_3^2) - (c_1 +c_3)f_2^2|}\,.
\end{align}
Compared to the convention in  Eq.~(\ref{eq:cseffective}),  we have dropped the index $\A$ from the coefficients $c_i$ as here there is only a single non-abelian gauge field.   We now impose a discrete $\mathbb{Z}_3$ symmetry under exchange of the throats on the Lagrangian, which implies $c_i \equiv c$ for all $i$,   and we take $c$ to be $\mathcal{O}(1)$ for simplicity. This exchange symmetry can be broken to a $\mathbb{Z}_2$  by a slightly differing length of the third throat (see e.g.~\cite{Law:2010pv}), such that we have  $L_{1,2} \equiv L$ whence $f_{1,2}\equiv f$, while $L_3=L+\varepsilon$ with $\varepsilon \ll k^{-1}$. The parameter $\varepsilon$ quantifies the $\mathbb{Z}_3$ breaking. Under this assumption, we can have both $f_{\A,i}^\prime$ with super-Planckian values.  We then compute the mass matrix and rotate to the mass basis $(a''_1,a''_2)$, with $a''_1$ being the state associated with the zero eigenvalue, which decouples. The potential for $a''_2$ is then given by Eq.~(\ref{AxionPotential}), with $1/f_{\textrm{eff}}^2 \equiv 1/f_{\A,1}^{\prime 2} + 1/f_{\A,2}^{\prime 2}$:
\be
f_{\rm eff} \approx \frac{f f_3 \sqrt{2 f^6 -3 f^4f_3^2 + f_3^6}}{ (f^2-f_3^2)^2}\, .
\ee
As expected from the combination of  $f_{\A,1}$ and $f_{\A,2}$, this can achieve trans-Planckian  values as a small denominator is obtained with $f_3\sim f\,(1-k \,\varepsilon)$.  The effective decay constant is then $f_{\rm eff} \sim f/(k\,\varepsilon)$.

  \bibliography{Bib}

\end{document}